\documentclass[fleqn,usenatbib]{mnras}
\usepackage{newtxtext,newtxmath}
\usepackage[T1]{fontenc}
\DeclareRobustCommand{\VAN}[3]{#2}
\let\VANthebibliography\thebibliography
\def\thebibliography{\DeclareRobustCommand{\VAN}[3]{##3}\VANthebibliography}
\usepackage{lscape}
\usepackage{graphicx}	
\usepackage{amsmath}	

\title[Massive Star Population in Sextans A]{Massive Star Population in the Sextans A Dwarf Galaxy from HST UV Photometry}

\author[R. Flores et al.]{
Roberto Flores,$^{1}$\thanks{E-mail: rflores\_7@yahoo.com}
E. Fantino,$^{1,2}$\thanks{E-mail: elena.fantino@ku.ac.ae}
G. Battaglia$^{3,4}$
and A. Aparicio$^{4,3}$
\\
$^{1}$Department of Aerospace Engineering, Khalifa University of Science and Technology, P.O. Box 127788, Abu Dhabi, United Arab Emirates\\
$^{2}$Polar Research Center, Khalifa University of Science and Technology, P.O. Box 127788, Abu Dhabi, United Arab Emirates\\
$^{3}$Instituto de Astrofisica de Canarias, Calle Via Lactea s/n 38206 La Laguna, Santa Cruz de Tenerife, Spain\\
$^{4}$Departamento de Astrofisica, Universidad de La Laguna, Avda. Astrofisico Francisco Sanchez 38205 La Laguna, Santa Cruz de Tenerife, Spain\\
}

\date{Accepted XXX. Received YYY; in original form ZZZ}

\pubyear{2025}

\begin{document}
\label{firstpage}
\pagerange{\pageref{firstpage}--\pageref{lastpage}}
\maketitle

\begin{abstract}
We build a catalog of massive (M>8~M$_\odot$) main sequence stars in the \mbox{metal-poor} ($\sim0.1$~Z$_\odot$) dwarf irregular galaxy Sextans A. HST WFC3 UV photometry in the 275 and 336 nm wideband filters is arranged in a Color-Magnitude Diagram (CMD), and overlaid on top of stellar evolutionary tracks from the MIST library. The star properties (mass, age, etc.) are computed with a Finite Element (FE) interpolation of the stellar tracks. The FE method, originally developed for solid mechanics problems, provides a general framework for interpolating fields inside domains of complex geometry. Besides the interpolated properties, the algorithm computes their gradients with respect to the photometry. These sensitivities provide a direct an efficient estimate of the associated uncertainties. Our catalog contains 655 stars, with the most massive one estimated at $58\pm11$~M$_\odot$. A comparison with a ground-based spectroscopic census of OB stars yields only 8 matches, evidencing the minimal overlap between both datasets. The mass estimates derived from the UV CMD and the spectral classification are in good agreement for the majority of O-type stars found in both datasets. Our catalog provides an extensive list of candidates for followup spectroscopic observation, which could improve our understanding of the early evolutionary stages of massive \mbox{low-metallicity} stars.
\end{abstract}


\section{Introduction}
\label{sec:intro}
Massive stars (M>8~M$_\odot$) play a major role in the evolution of galaxies. They are responsible for the vast majority of ionizing radiation, and their brightness dominates the total light output of star-forming regions. Furthermore, their powerful stellar winds and supernova explosions affect both star formation and chemical evolution of galaxies. Therefore, gaining insight on their properties is vital to understand the evolution of galaxies. Massive stars formed in \mbox{low-metallicity} environments are of special interest, as they are surrogates for the conditions of the early Universe. For a long time, the Small Magellanic Cloud (SMC, $\sim0.2$~Z$_\odot$) has been the prime target for studying metal-poor environments \citep{Lorenzo2022} due to its proximity to the Sun \citep[$60\pm7$~kpc,][]{Steer2017}. With the advent of modern instrumentation, considerable attention has been devoted to the dwarf irregular galaxy Sextans A. It is a very low metallicity object \citep[0.1~Z$_\odot$,][]{Lorenzo2022}, and its distance from the Sun \citep[1.34~Mpc,][]{Tammann2011} makes it possible to resolve individual stars. The massive stars in Sextans A can teach us about stellar evolution in conditions mimicking  the early Universe. According to \citet{Madau2014}, the chemical abundance of Sextans A is similar to the central regions of galaxy clusters at redshift $\sim2$.

\begin{figure}
	\includegraphics[width=\columnwidth]{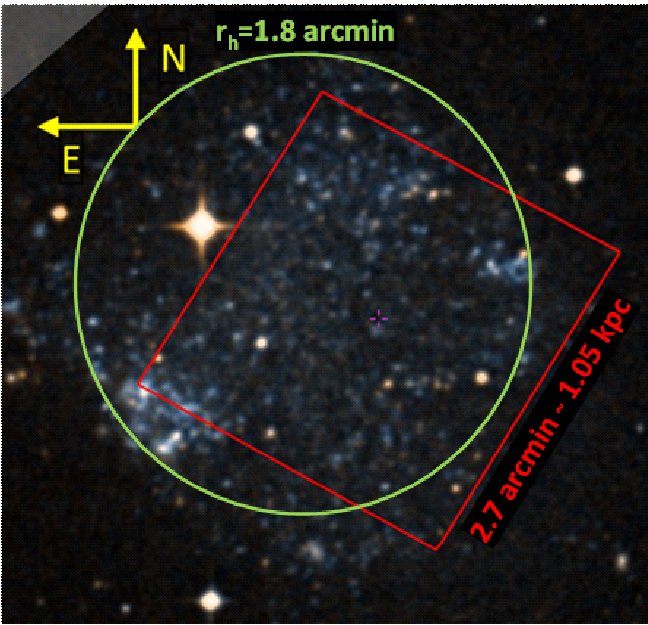}
	\caption{HST coverage of Sextans A (red square) and half-light circle (green circle). Background image credit: ESA/Aladin-Strasbourg Observatory.}
	\label{fig:sexa_coverage}
\end{figure}

Fritz Zwicky discovered Sextans A in 1942. It is located at RA $10^h 11^m 1^s.8$ Dec $-4^{\circ} 41' 34''$ \citep[J2000,][]{McConnachie2012}, with a 2D half-light radius $r_h=1.8$~arcmin \citep{Bellazzini2014}, equivalent to~700 pc (Fig.~\ref{fig:sexa_coverage}). The prevailing consensus is that Sextans A is not gravitationally bound to the local group \citep{vandenBergh1999}, and its motion is dominated by the Hubble flow \citep{McConnachie2012}. The galaxy is part of the Antlia-Sextans group, that also includes NGC 3109, Antlia, Sextans B and Leo P. The members of this association are tightly aligned in a 1~Mpc-long structure. This linear arrangement could be the result of a tidal interaction, but it could have formed in a thin cold cosmological filament instead \citep{Bellazzini2013}. Sextans A has a distinctive square shape in the visible spectrum. It is believed to be the result of an expanding shell of neutral hydrogen compressed by supernova explosions and, to a lesser extent, stellar winds from massive stars, that drives star formation \citep{vanDyk1998}.

Sextans A is a site of active star formation. The first in-depth study of its massive star population \citep{Aparicio1987} revealed multiple localized episodes of star formation spread over a 5x5~arcmin ($2.8 r_h$ across) region during the last 100~Ma\footnote{This document uses the symbol “a” for year, following the IAU style manual \citep{Wilkins1990}.}. According to \citet{Dolphin2003b}, after an initial stage of moderate activity, the galaxy became quiescent $\sim10$~Ga ago. Then, between 1 and 2~Ga ago, the star formation increased sharply, with another burst of activity in the last 100~Ma. Due to the ongoing star formation, Sextans A contains an abundance of young massive stars.

This work presents a methodology to estimate the properties of massive stars using UV CMDs combined with synthetic photometry derived from stellar evolution models. Hot stars are difficult to study in the visible range, because the CMD suffers from color degeneracy \citep{deMarchi1993}. The UV CMD alleviates  this issue, although  some level of degeneracy persists for the most massive stars (above 30~M$_\odot$). The stellar properties are computed interpolating the evolutionary tracks with a Finite Element (FE) discretization \citep{Zienkiewicz2013}. The FE method was originally developed to solve partial differential equations in the field of structural analysis. It provides a general framework for interpolating functions defined by discrete values, irrespective of the geometric complexity of the domain. Furthermore, it allows computing the gradient of those functions in an efficient and systematic way. The gradients yield a direct estimation of the uncertainty of the interpolated properties due to errors in the photometry.

The photometry was extracted from HST/WFC3 images in the 275 and 336~nm wide-band filters. Due to the superior resolving power of the space telescope \citep[FWHM $\sim80$~mas,][]{WFC3_IHB}, the effects of crowding in densely populated fields are less severe than for seeing-limited ground imaging. After correcting the apparent magnitudes for distance and extinction, the FE interpolation gives the estimated initial mass and age of the stars, together with their sensitivity to changes in flux for the two filters. The sensitivities, together with the estimated errors of the photometry, are used to compute the uncertainties in mass and age.

The procedure is very sensitive to errors in the color index, to the point that the typical width of the MS in the color direction is comparable to the uncertainties of the apparent magnitudes. To overcome this limitation, a reddening correction is applied based on a simple heuristic. It is assumed that the most likely position of the theoretical tracks relative to the observed stars places the maximum number of objects inside the MS. This corrected data yields a catalog of young massive stars that make excellent candidates for spectroscopic observations.

This document is structured as follows: Section~\ref{sec:material} characterizes the dataset and summarizes the assumed properties of Sextans A, the algorithm used to estimate stellar properties and their uncertainties, and the heuristic method to minimize errors in the color index; Sec.~~\ref{sec:results} presents our census of massive MS stars and its main features, and compares it to a ground-based spectroscopic catalog; finally, Sec.~\ref{sec:conclusions} summarizes the work and draws the main conclusions.

\section{Materials and methods}
\label{sec:material}
\subsection{HST dataset}
\label{sec:hst_data}
The images used to extract the photometry were captured during the GO-15275 campaign “Securing HST’s UV Legacy in the Local Volume: Probing Star Formation and the Interstellar Medium in Low Mass Galaxies”(PI Karoline M. Gilbert). Its objective is to complement existing visible and NIR  observations of 22 low-mass star-forming galaxies in the local universe (up to $\sim 3.8$~Mpc), to build a multiband panchromatic resolved stellar catalog. The first results of the  multiband dataset have been published recently \citep{Gilbert2025}, analyzing the quality of the photometry. However, the actual photometry has not been released, so this work has been carried independently with a different reduction pipeline.

\begin{figure}
\includegraphics[width=\columnwidth]{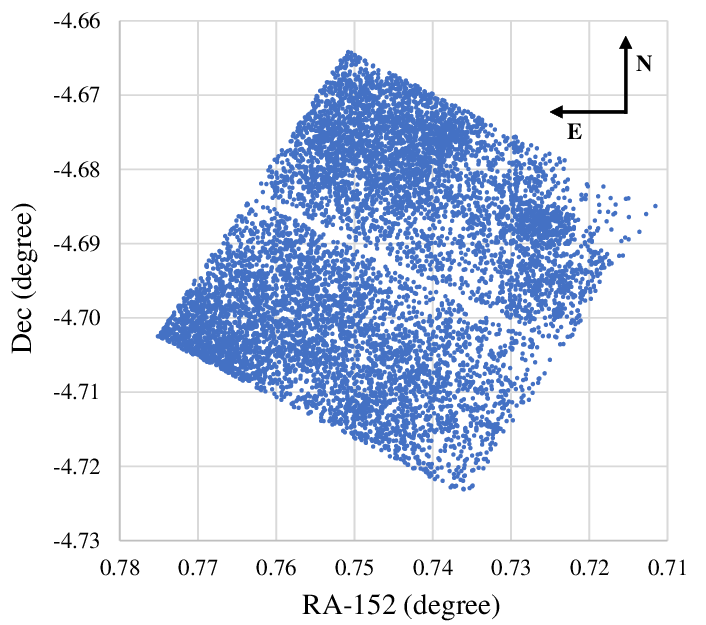}
\caption{Position of stars with valid UV photometry.}
\label{fig:hst_dataset}
\end{figure}

The data for Sextans A were obtained with a single pointing of the HST, using the 275 and 336~nm wideband filters of the WFC3/UVIS imager \citep{WFC3_IHB}. Four exposures were taken for each filter, for a total integration time of 2523~s at 275~nm and 2413~s at 336~nm. The instrument uses two 2k x 4k CCDs covering a field of view of 162 x 162~arcsec, with a scale of 0.0395~arcsec per pixel (Fig.~\ref{fig:sexa_coverage}). Because all the exposures correspond to one pointing of the telescope, there is a gap of 1.2~arcsec in the data, due to the separation between the CCDs (see Fig.~\ref{fig:hst_dataset}).

We started from pipeline-processed, charge-transfer-efficiency (CTE) corrected drizzled images (DRC products), generated by the standard WFC3/UVIS calibration pipeline (\texttt{calwf3} and \texttt{astrodrizzle}), which includes the appropriate pixel-area correction through the drizzling procedure \citep{WFC3_IHB,WFC3_DHB}.

Stellar photometry was carried out using point-spread function (PSF) fitting techniques as implemented in the DAOPHOT package within IRAF \citep{Stetson1987}. An empirical PSF was constructed independently for each filter from a set of bright, isolated stars uniformly distributed across the field. The resulting PSF models were then applied to the full stellar sample to measure instrumental PSF magnitudes.

Standard quality selection criteria of DAOPHOT were adopted to retain only reliable stellar measurements. In particular, objects with $|\texttt{sharpness}|>1$ or signal-to-noise ratio below 5 were removed to avoid spurious detections, extended sources, and poorly measured objects.

Instrumental PSF magnitudes were converted to aperture magnitudes following the standard photometric calibration procedures recommended by STScI for WFC3/UVIS data \citep{WFC3_IHB}. Aperture corrections were derived by comparing PSF magnitudes with aperture photometry measured within a radius of 10 pixels for the PSF stars. These magnitudes were then corrected to an infinite aperture using the tabulated encircled-energy corrections provided in the WFC3/UVIS photometric calibration documentation \citep{WFC3_EE}.

The final photometric calibration was performed using the official WFC3/UVIS zeropoints appropriate for each filter and for the epoch of the observations (March 2018), as provided by STScI. Zero-points were taken from the image headers and associated calibration tables, and all magnitudes were placed on the VEGAMAG system \citep{WFC3_IHB}. The final dataset contains a total of 9412 stars (Fig.~\ref{fig:hst_dataset}).

\subsection{Assumed properties for Sextans A}
\label{sec:sexa_prop}

\subsubsection{Extinction}
\label{sec:extinction}
The extinctions for the two UV passbands are derived from the visible color excess $E(B-V)_{SFD}$ using the coefficients in Table~\ref{tab:filters}. The subscript SFD denotes the galactic extinction map of \citet{Schlegel1998}, which is based on dust emission in the far infrared. The extinction coefficients were derived by \citet{Schlafly2011}. They recalibrated the SFD map against SDSS (Sloan Digital Sky Survey) data. The coefficients assume a \citet{Fitzpatrick1999} law with $R_V=A_V/E(B-V)=3.1$.

\begin{table}
	\centering
	\caption{Filter throughput-weighted mean wavelengths ($\lambda_{eff}$) and extinction coefficients. Fitzpatrick's law with $R_V=0.31$ \citep{Schlafly2011}.}
	\label{tab:filters}
	\begin{tabular}{lcc}
		\hline
		Bandpass (b) & $\lambda_{eff}$ (nm) & $A_b/E(B-V)_{SFD}$\\
		\hline
		HST WFC3 F275W & 274.25 & 5.487\\
		HST WFC3 F336W & 336.64 & 4.453\\
		\hline
	\end{tabular}
\end{table}

For Sextans A, the foreground (fg) color excess is $E(B-V)_{SFD,fg}=0.044$. Note that, for star-forming galaxies such as Sextans A, measurements of the Galactic dust emission may be contaminated by the internal emission \citep{Choi2025}. Because the dust attenuation curve is different in \mbox{low-metallicity} environments, the accuracy of the extinction determination is degraded.

In an effort to improve the reddening estimates of the OB population in dwarf galaxies with recent star formation, \citet{Massey2007} made comparisons of their CMDs with those of M31, M33, LMC and SMC. Their result for Sextans A is $E(B-V)=0.05 \pm 0.05$. While this value is compatible with SFD, it highlights the substantial uncertainty. Errors in the color index have a large potential effect on the identification of properties using the UV CMD, due to the strong sensitivity to this parameter (typical variation of color index across the MS: $\sim0.1$~mag). Therefore, a reddening calibration becomes necessary in order to extract useful data. This is explicated in Sec. \ref{sec:red_cal}.

It is often assumed that the internal extinction of dwarf irregular galaxies is very small \citep{Lorenzo2025}. For example, \citet{Tammann2011} estimated the internal (int) reddening of Sextans A during a calibration of the period-luminosity and period-color relations of Cepheids. Their result was $E(B-V)_{int}=-0.012 \pm 0.012$, which is compatible with no internal color excess. \citet{Garcia2019} compared observed photometry with intrinsic colors derived from spectral classification for massive stars in the South of Sextans A. They found an uneven reddening, peaking at $E(B-V)_{int}=0.2$. This is further supported by the spectroscopic catalog of \citet{Lorenzo2022}, where the position of OB stars in the CMD did not match expectation based on the spectral type. Recently, \citet{Lorenzo2025} built a calibration of intrinsic colors as a function of the spectral type, using stellar atmosphere models. Comparing the observed vs. expected color excess revealed that, while the internal reddening is small for the majority of stars, it is significant for some of them. Furthermore, the variation of internal extinction occurs over small spatial scales, and does not correlate well with the HI column density map. This uncertainty affecting the internal color excess further stresses the need for a reddening calibration.

\subsubsection{Distance modulus}
\label{sec:dis_mod}
Table~\ref{tab:dis_mod} presents a compilation of distance moduli from the literature. They all agree to better than 0.1~mag, considering their confidence intervals. For this work, the true distance modulus $\mu^0 = (m - M)^0 = 25.63 \pm 0.03$ from \citet{Tammann2011}, derived from Cepheid measurements, has been adopted. It corresponds to a distance of $1.34 \pm 0.02$~Mpc. The reference also includes an alternate determination, lumping together data for Sextans A and B. It gives $\mu^0 = 25.60 \pm 0.03$ for the pair, which is compatible with the previous value.

\begin{table}
	\centering
	\caption{True distance moduli from the literature.}
	\label{tab:dis_mod}
	\begin{tabular}{lc}
		\hline
		Source & $\mu^0$\\
		\hline
		\citet{Aparicio1987} & $25.60 \pm 0.30$\\
		\citet{Piotto1994} & $25.71 \pm 0.20$\\
		\citet{Sakai1996} & $25.87 \pm 0.15$\\
		\citet{Dolphin2003a} & $25.66 \pm 0.03$\\
		\citet{Tammann2011} & $25.63 \pm 0.03$\\
		\citet{Tammann2011} (Sex. A+B) & $25.60 \pm 0.03$\\
		\citet{McConnachie2012} & $25.78 \pm 0.08$\\
		\hline
	\end{tabular}
\end{table}

The expected impact of the distance uncertainty on the stellar properties is limited. Contrary to the color index, the sensitivity to errors in absolute magnitude is moderate (over the range of masses considered, the absolute magnitude varies by more than 6~mag).

\subsubsection{Metallicity}
\label{sec:metal}
To compute the stellar evolution tracks, an estimate of the metallicity is required. This work uses the MIST library \citep{Choi2016,Dotter2016}, which assumes a protosolar metallicity Z$_\odot = 0.0142$ \citep{Asplund2009}. The library takes a value of [Fe/H] as input, and scales all the element abundances from the solar values. Therefore, [Fe/H]=$-1$ is effectively equivalent to $Z=0.1$~Z$_\odot$. Table~\ref{tab:metal} contains a sample of metallicity estimates from the literature. For references that give the (O/H) ratio, an equivalent [Fe/H] has been computed assuming the abundances of all elements scale from the solar composition, with $12 + \log_{10} (\text{O/H})_\odot = 8.69$ \citep{Asplund2009}.

\begin{table}
	\centering
	\caption{Several Sextans A metallicity estimates from the literature.}
	\label{tab:metal}
	\begin{tabular}{lcc}
		\hline
		Source & $12 + \log_{10} (\text{O/H})$ & [Fe/H]\\
		\hline
		\citet{Sakai2004} & 7.49 & -1.2\\
		\citet{Lorenzo2022} & - & -1.0\\
		\citet{Dolphin2003a} & 7.6 & -1.1\\
		\citet{Skillman1989} & 7.49 & -1.2\\
		\citet{Pilyugin2001} & 7.71 & -0.98\\
		\citet{Kniazev2005} & $7.54 \pm 0.06$ & $-1.15 \pm 0.06$\\
		\citet{Kaufer2004} & - & $-1.03 \pm 0.16$\\
		\hline
	\end{tabular}
\end{table}

A value $Z=0.1$~Z$_\odot$ has been adopted, following \citet{Lorenzo2022}.

\subsection{Extraction of stellar properties from the CMD}
\label{sec:extraction}

\subsubsection{Preparation of stellar tracks}
\label{sec:prep_track}
The HST UV photometry contains the apparent magnitudes in the 275 nm and 336~nm filters of the WFC3, measured in the Vega system ($F275$ \& $F336$ hereafter). These values, once corrected for extinction and distance, yield the absolute magnitudes, which are  arranged in a CMD. For the sake of brevity, the two coordinates in the diagram shall be denoted as $x$ and $y$ henceforth, with
\begin{equation}
	x = F275-F336-(A_{F275}-A_{F336}) \; , \; y = F336-A_{F336}-\mu^0.
	\label{eq:y}
\end{equation}
When appropriate, a point in the diagram shall be written as a coordinate vector $\bf x$.

To estimate the properties of a star, its position on the CMD is compared against synthetic photometry data taken from the MIST stellar evolution library \citep{Choi2016,Dotter2016}. Based on the assumed properties for Sextans A, the stellar tracks were generated for [Fe/H]=$-1$. The rotation rate of stars during the Zero-Age MS (ZAMS) has a substantial influence on their subsequent evolution. For example, it increases the MS lifetime through enhanced mixing, and enhances mass loss from stellar winds \citep{Maeder2000}. The library offers two initial rotation rates, 0 and $V/V_{crit}=0.4$, where $V$ is the equatorial tangential velocity, and $V_{crit}$ is the value of $V$ for which the centrifugal force balances gravity. \citet{Huang2010} found that, in B stars, the $V/V_{crit}$ probability density peaks at 0.49 (although, for the most massive objects, the fraction of slow rotators increases). Thus, $V/V_{crit}=0.4$ has been selected. The initial mass range for the tracks covers from 4 to 100~M$_\odot$. The lower limit is below the minimum mass in the catalog (8 M$_\odot$), to ensure that it has no effect on the results. The mass increment between tracks is 0.0125~dex below 16.9~M$_\odot$, and 0.025~dex above. Larger steps in the upper mass range help to maintain a reasonable spacing between adjacent tracks.

While the stellar tracks include both the MS and Core Helium Burning (CHeB) stages, the latter has not been considered for the analysis. The reason is the lack of reliability of the stellar evolution models for \mbox{low-metallicity} stars. When the MS and CHeB tracks are plotted on the CMD, there is a wide gap between them in the color direction ($\sim0.3$~mag). This region corresponds to the beginning of the Red Giant Branch (RGB) phase, right after the Terminal-Age MS (TAMS). Due to its relatively short duration, very few stars should be found in this part of the CMD. However, the observational data paints a vastly different picture, with a substantial number of objects falling inside the gap. This issue has been reported in the literature. \citet{Tang2014} found that, in star-forming dwarf galaxies with metallicity below the LMC, the theoretical tip of the blue loop is too red compared with the observed blue He-burning massive stars. The evolutionary tracks could only be reconciled with the observations through enhanced overshooting at the base of the convective envelope during the first dredge-up. The mixing scales required are considerably larger than those in common use. Likewise, a study of dwarf galaxies by \citet{Cignoni2018}, found that the wide color gap between the MS and CHeB tracks does not match the observations. According to \citet{Choi2025}, addressing the issue through changes of the  chemical composition requires an unrealistic metallicity, too low by an order of magnitude. Furthermore, it causes the red end of the CHeB tracks to diverge from the observations. Due to these shortcomings, only the MS has been considered.

\begin{figure}
	\includegraphics[width=\columnwidth]{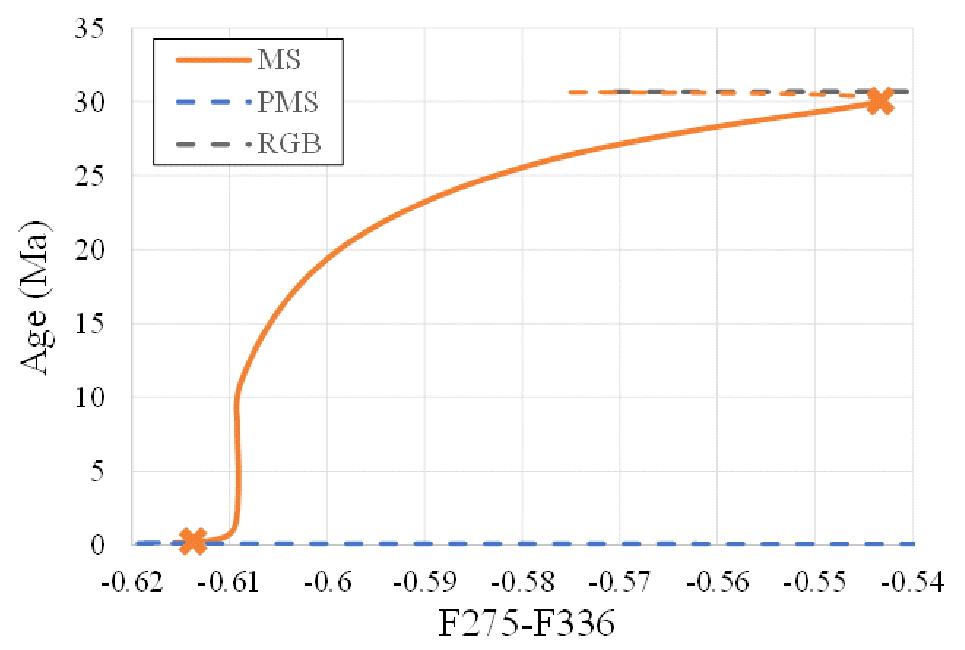}
	\caption{Color-age diagram highlighting the PMS (blue), MS (orange) and RGB (black) stages. Dashed lines mark the removed sections, segment between orange crosses is retained. Plot shows a track computed for 8.95~M$_\odot$, [Fe/H]=$-1$, $V/V_{crit}=0.4$.}
	\label{fig:track_prep}
\end{figure}

There is a color degeneracy at the end of the MS, due to a fast motion towards the blue side of the CMD just before the TAMS. An example for a 9~M$_\odot$ star is shown in Fig.~\ref{fig:track_prep}. To avoid intersections between neighboring tracks, the MS is cropped at the point where the derivative of the color with respect to time changes sign. The orange crosses in the figure signal the segment of the MS that is retained. The section removed represents less than 2 per cent of the total MS duration. Its expected effect on the estimated properties is therefore limited.

\begin{figure}
	\includegraphics[width=\columnwidth]{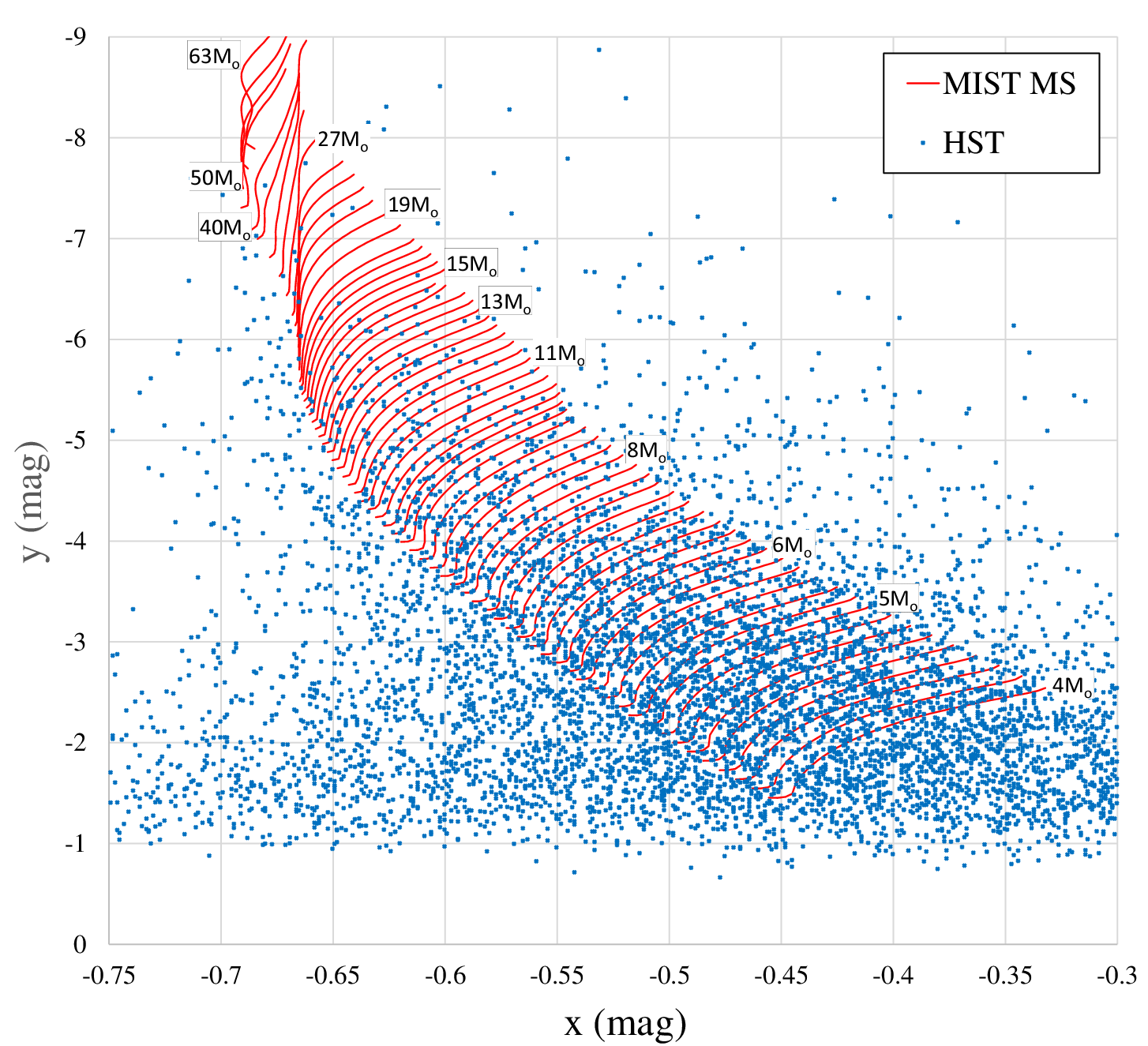}
	\caption{CMD with HST data (blue markers) and MS tracks (red lines). [Fe/H]=$-1$, $V/V_{crit}=0.4$, $\mu^0=25.63$.}
	\label{fig:cmd_tracks}
\end{figure}

Finally, the clipped tracks are regularized by sampling them at equally spaced intervals in the CMD space. This accelerates the search for close neighbors, and helps control the distortion of the interpolation domains. The CMD (before reddening calibration), with the MS tracks superimposed, is shown in Fig~\ref{fig:cmd_tracks}.

\subsubsection{Finite element interpolation}
\label{sec:fe_int}
To assign properties to a point of the CMD, a FE interpolation \citep{Zienkiewicz2013} of the stellar track data has been used. Given a set of nodes ${\bf x}_i$, where a sought property $\phi$ is known, the value at a generic point $\bf x$ is computed as
\begin{equation}
	\phi({\bf x}) = \sum\limits_{i=1}^n N_i({\bf x}) \, \phi _i,
	\label{eq:fe_int}
\end{equation}
where $n$ is the number of nodes in the interpolation domain (known as “element”). The functions $N_i(\bf{x})$ are called shape functions. They have the property
\begin{equation}
	N_i({\bf x}_j) = \delta_{ij},
	\label{eq:n_ij}
\end{equation}
with $\delta_{ij}$ denoting the Kronecker delta function:
\begin{equation}
	\delta _{ij} = \left\{ {\begin{array}{c}
		0 \quad \text{for} \quad i \ne j\\
		1 \quad \text{for} \quad i = j
	\end{array}} \right.
\end{equation}
Thus, each shape function takes the value 1 at its own node, and 0 at the others. Furthermore, it is usually required that, for any point $\bf x$, 
\begin{equation}
	\sum\limits_{i = 1}^n N_i({\bf x}) = 1.
	\label{eq:sum_ni}
\end{equation}
The condition (\ref{eq:sum_ni}) ensures that the interpolation is consistent, meaning a constant field is reproduced exactly.

For first order interpolation (deemed sufficient given the uncertainties of the photometry and stellar evolution modeling), the most common choices for $n$ are 3 (triangles) and 4 (quadrilaterals). Quadrilaterals, when they are not distorted (i.e., when they are close to rectangular), are usually preferred because they yield a better approximation of the function gradient. However, given that some adjacent tracks have substantially different shapes, it is extremely difficult to completely avoid distortion. For this reason, triangular elements have been chosen, as they offer increased robustness. Even if the interpolation quality is not as good, taking into account the uncertainties of the data, they are an effective option. Furthermore, the mathematical apparatus becomes simpler and faster to evaluate.

\begin{figure}
	\includegraphics[width=\columnwidth]{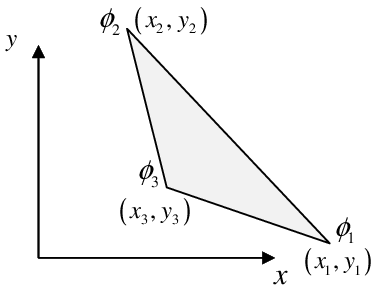}
	\caption{Triangular element in global (CMD) coordinates.}
	\label{fig:tria_glob}
\end{figure}

\begin{figure}
	\includegraphics[width=\columnwidth]{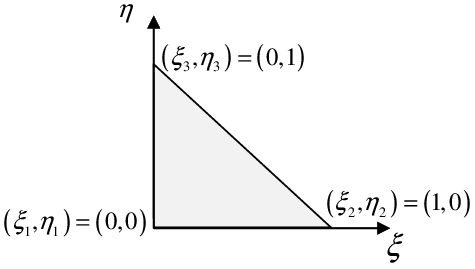}
	\caption{Local element reference frame.}
	\label{fig:tria_loc}
\end{figure}

While it is possible to find a closed expression for the shape functions of the first-order triangle, it is cumbersome and unintuitive. Instead, it is simpler to switch from the CMD coordinates (Fig.~\ref{fig:tria_glob}) to a local element reference frame $(\xi, \eta)$, where the geometry and associated algebra become straightforward (Fig.~\ref{fig:tria_loc}).

The expressions of the shape functions in the local coordinate system are trivial. To simplify the derivation, the functions will be arranged in matrix form:
\begin{equation}
	{\bf N}^T = \begin{bmatrix}
		N_1\\
		N_2\\
		N_3
	\end{bmatrix} = \begin{bmatrix}
		{1 - \xi  - \eta }\\
		\xi \\
		\eta 
	\end{bmatrix}.
\end{equation}
It is easy to verify that the conditions (\ref{eq:n_ij}) and (\ref{eq:sum_ni}) are fulfilled. It is possible to convert between global and local coordinates using the shape functions themselves. This is called an isoparametric transform:
\begin{equation}
	{\bf x}(\xi, \eta) = \sum\limits_{i=1}^n N_i(\xi, \eta) \, {\bf x}_i,
\end{equation}
which can be written in matrix form as
\begin{equation}
	\begin{bmatrix}
		x & y
	\end{bmatrix} = {\bf N} \, {\bf X} = {\bf N} \, \begin{bmatrix}
		x_1 & y_1\\
		x_2 & y_2\\
		x_3 & y_3
	\end{bmatrix}.
	\label{eq:iso_param}
\end{equation}
Equation~\ref{eq:iso_param} gives $(x, y)$ as an explicit function of $(\xi, \eta)$. To invert the transformation, start by computing the gradients of the shape functions in local coordinates
\begin{equation}
	{\bf G} = \begin{bmatrix}
		N_{1,\xi} & N_{2,\xi} & N_{3,\xi}\\
		N_{1,\eta} & N_{2,\eta} & N_{3,\eta}
	\end{bmatrix} = \begin{bmatrix}
		-1 & 1 & 0\\
		-1 & 0 & 1
	\end{bmatrix},
	\label{eq:gra_loc}
\end{equation}
where the partial derivative of $N_1$ with respect to $\xi$ is denoted as $	N_{1,\xi}$, and so on. Note that $\bf G$ is a constant matrix. In structural analysis, the first order triangular element is often disfavored because it corresponds to a constant strain field. This limits the accuracy of the stress calculations. However, because we use the gradient only to estimate the sensitivities of the interpolated properties to errors in photometry, the constant gradient approximation is deemed acceptable.
Combining Eqs.~\ref{eq:iso_param} and \ref{eq:gra_loc} yields the Jacobian of the transform
\begin{equation}
	{\bf J} = \begin{bmatrix}
		\partial x / \partial \xi & \partial y / \partial \xi\\
		\partial x / \partial \eta & \partial y / \partial \eta\\
	\end{bmatrix} = {\bf G} \, {\bf X},
	\label{eq:jaco}
\end{equation}
which is also constant. Because the Jacobian is constant, and node 1 has local coordinates (0,0), the expression of the local coordinates as a function of the global ones is straightforward:
\begin{equation}
	\begin{bmatrix}
		x\\
		y
	\end{bmatrix} = \begin{bmatrix}
	x_1\\
	y_1
	\end{bmatrix} + {\bf J}^T \begin{bmatrix}
	\xi\\
	\eta
	\end{bmatrix} \rightarrow \begin{bmatrix}
	\xi\\
	\eta
	\end{bmatrix} = \left( {\bf J}^{-1} \right)^T \begin{bmatrix}
	x - x_1\\
	y - y_1
	\end{bmatrix}.
\end{equation}
From Eqs.~\ref{eq:fe_int} and \ref{eq:gra_loc}, the gradient of the function $\phi$ in local coordinates is given by
\begin{equation}
	\begin{bmatrix}
		\partial \phi / \partial \xi\\
		\partial \phi / \partial \eta
	\end{bmatrix} =  {\bf G} \, \begin{bmatrix}
		\phi_1\\
		\phi_2\\
		\phi_3
	\end{bmatrix} = {\bf G} \, {\bf \Phi}.
	\label{eq:gra_phi_loc}
\end{equation}
The relation between the gradients of $\phi$ in local and global coordinates is
\begin{equation}
	\begin{bmatrix}
		\partial \phi / \partial \xi\\
		\partial \phi / \partial \eta
	\end{bmatrix} = {\bf J} \, \begin{bmatrix}
		\partial \phi / \partial x\\
		\partial \phi / \partial y
	\end{bmatrix}.
	\label{eq:rel_gra}
\end{equation}
Inverting Eq.~\ref{eq:rel_gra} and substituting Eq.~\ref{eq:gra_phi_loc}, the gradient in $(x,y)$ coordinates (i.e., in the CMD plane) becomes
\begin{equation}
	\nabla \phi = \begin{bmatrix}
		\partial \phi / \partial x\\
		\partial \phi / \partial y
	\end{bmatrix} = {\bf J}^{-1} \begin{bmatrix}
		\partial \phi / \partial \xi\\
		\partial \phi / \partial \eta
	\end{bmatrix} = {\bf J}^{-1} {\bf G} \, {\bf \Phi}.
\end{equation}
Thus, the matrix ${\bf J}^{-1} {\bf G}$ contains the weights that determine the gradient of an arbitrary function from its nodal values.

\subsubsection{Selection of the interpolation domain}
\label{sec:sel_dom}
Given a point $\bf p$ in the CMD (see Fig.~\ref{fig:cand_els}), the selection of the nodes that best interpolate the tracks follows these steps:

\begin{enumerate}
\item Find the closest point to $\bf p$ in the track set\footnote{Internally, the algorithm scales the $x$ and $y$ coordinates to have comparable ranges of variation for both. The CMD is usually plotted with a much larger scale for the $y$ axis. If this were not accounted for, the search direction for the closest point would be always quasi-horizontal, which is counterintuitive. By scaling the coordinates, the search algorithm behaves closer to what a human would do.}. Let the coordinates of this point be ${\bf x}_{j,k}$. Index $j \in \{1,...,n_{ppt}\}$   indicates the position along the track and $k \in \{1,...,n_t\}$ identifies the track. The number of tracks used for the calculations is $n_t = 66$. Due to the regularization process, the number of points per track $n_{ppt}$ is the same for all tracks. Furthermore, the points of each track are always arranged in increasing values of color, such that $x_{j-1,k} < x_{j,k} < x_{j + 1,k}$.

\item Build the four candidate elements $\{A,B,C,D\}$ that surround ${\bf x}_{j,k}$ as shown in Table~\ref{tab:cand_els}. Note that when ${\bf x}_{j,k}$ lies at the edge of the track set (i.e., $i=1, n_{ppt}$ or $j=1, n_t$), some of the combinations in Table~\ref{tab:cand_els} are not feasible.

\begin{table}
	\centering
	\caption{Node choices for interpolation.}
	\label{tab:cand_els}
	\begin{tabular}{lccc}
		\hline
		Element & Node 1 & Node 2 & Node 3\\
		\hline
		A & ${\bf x}_{j,k}$ & ${\bf x}_{j+1,k}$ & ${\bf x}_{j,k+1}$\\
		B & ${\bf x}_{j,k}$ & ${\bf x}_{j,k+1}$ & ${\bf x}_{j-1,k}$\\
		C & ${\bf x}_{j,k}$ & ${\bf x}_{j-1,k}$ & ${\bf x}_{j,k-1}$\\
		D & ${\bf x}_{j,k}$ & ${\bf x}_{j,k-1}$ & ${\bf x}_{j+1,k}$\\
		\hline
	\end{tabular}
\end{table}

\begin{figure}
	\includegraphics[width=\columnwidth]{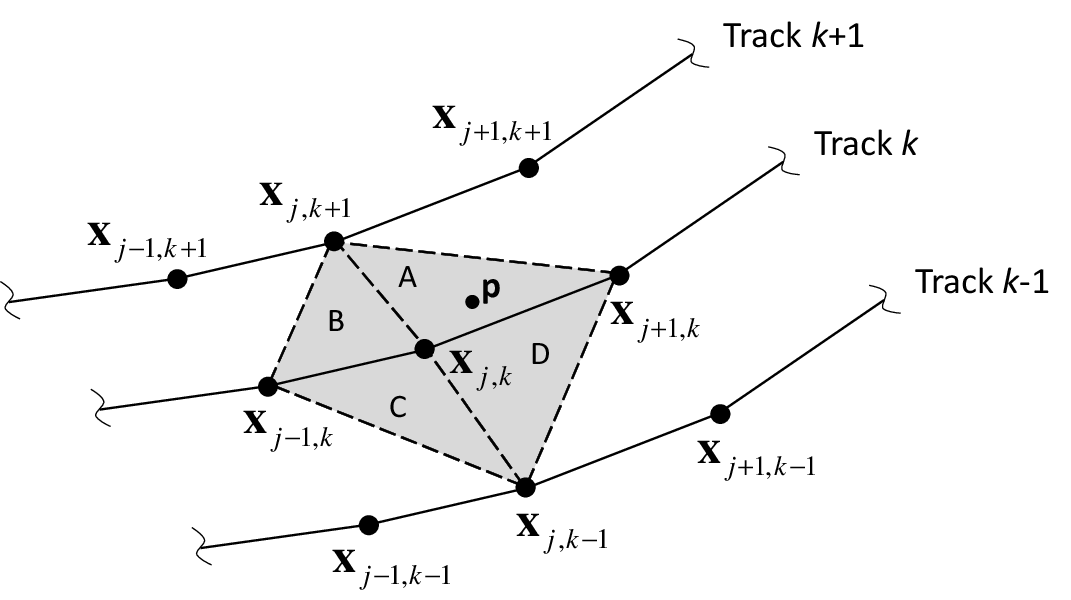}
	\caption{Selection of interpolation nodes.}
	\label{fig:cand_els}
\end{figure}

\item Retain the element for which both local coordinates of $\bf p$ are positive (i.e., $\bf p$ lies in the first quadrant of the $(\xi,\eta)$ plane, where the interior of the element is). This way, the value can be interpolated instead of extrapolated, improving accuracy. For the example in Fig.~\ref{fig:cand_els}, element A would be selected. 
\end{enumerate}

\subsubsection{Points falling outside the track set}
\label{sec:poi_out}
There are points in the CMD that fall outside the envelope of the track set, but with uncertainties in the photometry compatible with the actual value being inside. Their properties have been estimated based only on the $y$ coordinate of the CMD. Fig.~\ref{fig:extra_out} illustrates the case where the point lies to the left of the track set. The steps are:

\begin{figure}
	\includegraphics[width=\columnwidth]{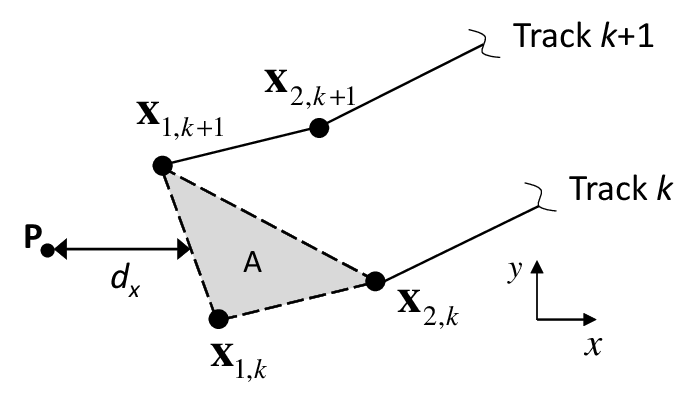}
	\caption{Extrapolation for points outside the track envelope.}
	\label{fig:extra_out}
\end{figure}

\begin{enumerate}
\item Find the nodes on the left edge of the track set $(j=1)$ which lie immediately above $(k+1)$ and below $(k)$ of point $\bf p$.

\item Compute the interpolation weigh
\begin{equation}
	\theta  = \frac{y_P - y_{1,k}} {y_{1,k+1} - y_{1,k}} \in [0,1].
\end{equation}

\item Retain the point only if the distance to the track envelope ($d_x$) is less than a threshold value $d_{\max}$:
\begin{equation}
	d_x = \theta \, x_{1,k+1} + (1-\theta) \, x_{1,k} - x_P < d_{\max}.
\end{equation}
For each object, the maximum distance is proportional to its color uncertainty ($\sigma_x$)
\begin{equation}
	d_{\max} = \beta \, \sigma _x,
\end{equation}
where $\beta$ is a proportionality constant set by the user. For example, $\beta=3$ corresponds to a 99.7 per cent confidence interval. In principle, the standard deviation of color would be evaluated as 
\begin{equation}
\sigma _x^2 = \sigma _{F275}^2 + \sigma _{F336}^2 - 2 \, \text{cov}(F275,F336).
\end{equation}
However, the covariance of the apparent magnitudes is not known. Instead, $\sigma_x$ has been estimated as \citep{Aparicio1995}
\begin{equation}
	\sigma _x \approx \max (\sigma _{F275}, \sigma _{F336}).
\end{equation}

\item The properties of the star are interpolated using
\begin{equation}
	\phi _x = \theta \, \phi _{1,k+1} + (1-\theta) \, \phi_{1,k}.
\end{equation}

\item The sensitivities to the photometry (i.e., the gradients of the properties) are taken from element A, as shown in Fig.~\ref{fig:extra_out}.
\end{enumerate}

\subsubsection{Estimation of uncertainties}
\label{sec:est_unc}
The uncertainty on the interpolated properties is derived from the error estimate of the photometry and the local sensitivity of the properties to the CMD coordinates. For any interpolated quantity $\phi$, its variation with respect to the photometry is
\begin{equation}
\delta \phi  \simeq \frac{\partial \phi}{\partial x} \delta x + \frac{\partial \phi}{\partial y} \delta y = \left(\frac{\partial \phi}{\partial x} + \frac{\partial \phi}{\partial y} \right) \delta F275 - \frac{\partial \phi}{\partial x} \delta F336.
\end{equation}
To keep the notation compact, the sensitivities shall be denoted as
\begin{equation}
	a_\phi = \frac{\partial \phi}{\partial x} + \frac{\partial \phi}{\partial y} \quad , \quad b_\phi =  - \frac{\partial \phi}{\partial x}.
\end{equation}
Using the sensitivities and the uncertainties of the photometry:
\begin{equation}
	\sigma_\phi^2 = a_\phi^2 \sigma_{F275}^2 + b_\phi^2 \sigma_{F336}^2 + 2 a_\phi b_\phi \text{cov}(F275,F336).
\end{equation}
Proceeding as in the case of the color, it has been assumed that
\begin{equation}
	\sigma_\phi \approx \max \left( \left| a_\phi \right| \sigma_{F275} \, , \, \left| b_\phi \right| \sigma_{F336} \right).
	\label{eq:uncer}
\end{equation}
In regions of color degeneracy or where the stellar tracks are highly distorted, the gradient of the properties experiences large changes over small distances. This occurs, for example, near $(x,y)$ coordinates (-0.67,-6.4) and (-0.69,-7.9) of the CMD (see Fig.~\ref{fig:cmd_tracks}). In those areas, Eq.~\ref{eq:uncer} can grossly overestimate the uncertainty. The occurrence is easy to detect, because it gives rise to mass error estimates on the order of 100 per cent or higher. The issue is mitigated switching to an alternate Monte Carlo algorithm. Samples of synthetic photometry are generated adding normally-distributed perturbations to the measured fluxes:
\begin{equation}
	\begin{array}{l}
		F275_{syn} = F275 + \delta_{F275} \; , \; F336_{syn} = F336 + \delta_{F336},\\
		\delta _{F275} \sim N(0,\sigma _{F275}) \; , \; \delta_{F336} \sim N(0,\sigma_{F336}).
	\end{array}
	\label{eq:mon_car}
\end{equation}
Then, the uncertainties are estimated repeating the identification process for a set of 50 synthetic samples and computing the variance of the stellar properties. Compared with the gradient-based approach, this algorithm is quite inefficient. However, it is only required for a small number of objects ($\sim10$), so the overall performance penalty is small. Note also that the Monte Carlo method yields less reliable error estimates for objects that fall outside of the track envelope. Their properties are computed from the $y$ coordinate only, so they are less sensitive to photometric errors. Thus, the gradient-based approach is the preferred choice.

\subsection{Limiting magnitude and photometric quality}
\label{sec:lim_mag}

\begin{figure}
	\includegraphics[width=\columnwidth]{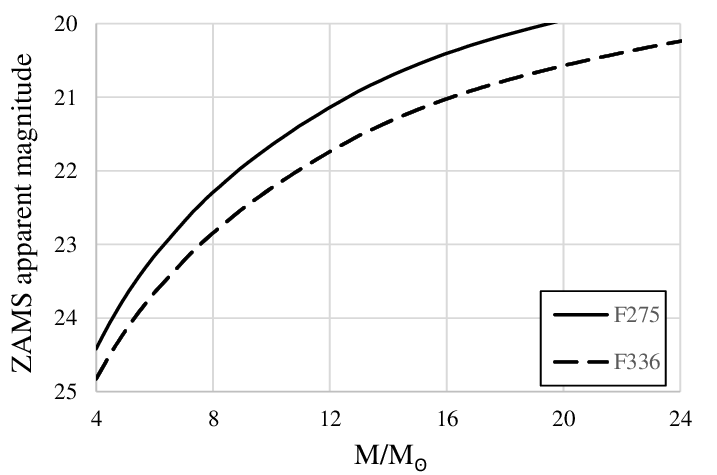}
	\caption{ZAMS apparent magnitude in 275 (solid line) and 336~nm (dashed line) filters vs. initial mass. [Fe/H]=$-1$, $V/V_{crit} = 0.4$, $E(B-V) = 0.044$, $\mu^0 = 25.63$.}
	\label{fig:zams_mag}
\end{figure}

Because the goal is to obtain a catalog of stars above 8~M$_\odot$, the limiting magnitude for the star count must be chosen accordingly. The apparent magnitude at ZAMS (the faintest point of the MS tracks, see Fig.~\ref{fig:cmd_tracks}) as a function of the initial mass is shown in Fig.~\ref{fig:zams_mag}. The plot corresponds to the foreground reddening measured by \citet{Schlegel1998}. Stars above 8~M$_\odot$ are brighter than 22.3~mag in $F275$ and 22.8~mag in $F336$. To ensure that all massive stars are included in the reddening calibration, the limit magnitude has been set at 23.5. This guarantees that, irrespective of the final value adopted for the extinctions, all stars heavier than 8~M$_\odot$ are included in the sample.

\citet{Gilbert2025}, using injection of artificial stars and determining the recovered fraction, estimate a 90 per cent recovery threshold of 25.0~mag for $F275$ and 25.7~mag for $F336$. This is over two magnitudes fainter than the limit selected for the reddening calibration. Therefore, photometric completeness is not an issue. 

\begin{figure}
	\includegraphics[width=\columnwidth]{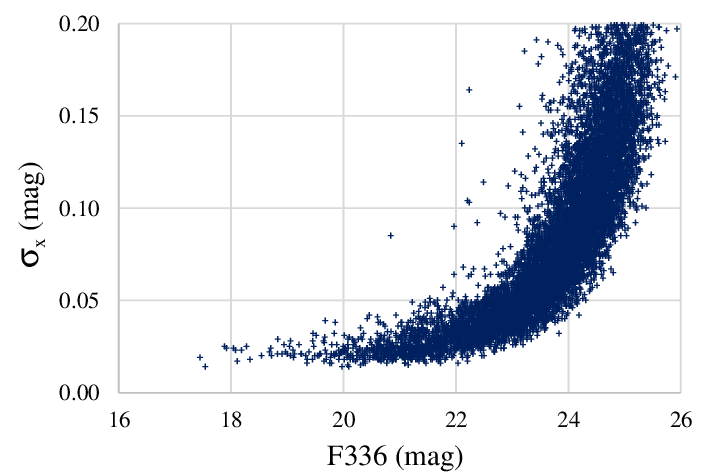}
	\caption{Color uncertainty vs. F275 apparent magnitude.}
	\label{fig:sx_f336}
\end{figure}

For a limiting $F336$ magnitude of 23.5, Fig.~\ref{fig:sx_f336} indicates that the typical color uncertainty of the photometry is below 0.08~mag. Furthermore, for massive stars ($F336 < 22.8$~mag), $\sigma_x$ is less than 0.05~mag.

\subsection{Reddening calibration}
\label{sec:red_cal}
As shown in Fig.~\ref{fig:cmd_tracks}, the width of the main sequence in the $x$ direction of the CMD is narrow. It decreases from around 0.1~mag at 6~M$_\odot$ to 0.05~mag at 19~M$_\odot$. This is comparable to the uncertainty of the color excess measured by \citet{Massey2007} (see Sect.~\ref{sec:extinction}). Therefore, color index errors can have a serious impact in the stellar properties derived from the CMD. The internal reddening is a major source of concern, because massive stars can have a large impact in their environment due to their strong stellar winds \citep{Lorenzo2022}. Moreover, they may form in compact pockets of increased gas density \citep{Lorenzo2025}, giving rise to extinction variations over small spatial scales.

Using only the $F275$ and $F336$ fluxes it is not possible to estimate the internal extinction for individual stars. However, given the large impact that color errors have on the estimated properties, a correction is required to account for internal extinction. A global reddening correction based on a simple heuristic has been adopted. Because stars spend most of their lives inside the main sequence, it is expected that the vast majority of the objects observed lie inside the MS of the CMD. It is likely that spurious changes of color result in more stars falling outside the expected region. Thus, it is plausible that, when the relative position of the tracks and stars in the CMD is correct, the number of objects found inside the MS is close to maximum. Taking $E(B-V)$ as independent variable and computing the UV extinctions with the coefficients of Table~\ref{tab:filters}, the color excess that yields the best alignment between the stellar tracks and the observed stars can be found.

The procedure also mitigates other adverse effects in the construction of the CMD, such as zero-point errors of the photometry and inacuracies of the stellar evolution models. It must be clear that the correction does not provide an estimation of the actual reddening (foreground plus internal), its only purpose is to place the observed stars in the most likely position relative to the stellar tracks.

\begin{figure}
	\includegraphics[width=\columnwidth]{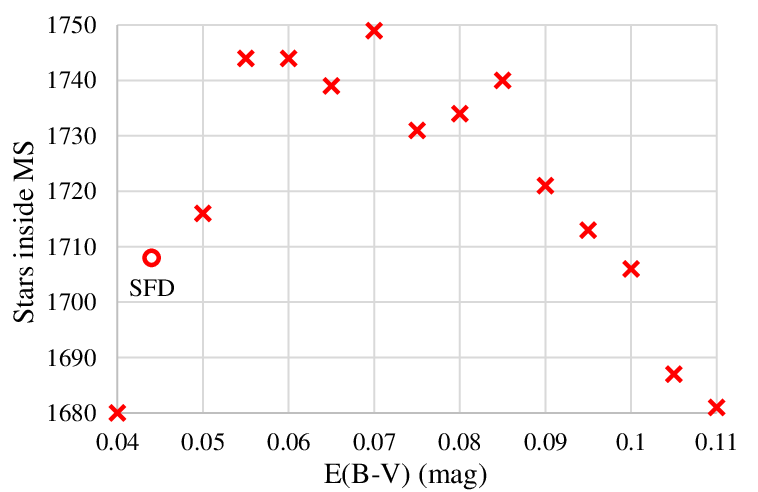}
	\caption{Number of stars inside MS tracks vs. $E(B-V)$. Circular marker (SFD) denotes the reddening estimated by \citet{Schlegel1998}. [Fe/H]=$-1$, $V/V_{crit} = 0.4$, $\mu^0 = 25.63$, $F336<23.5$~mag.}
	\label{fig:red_cal}
\end{figure}

We applied the interpolation methodology presented in Sec.~\ref{sec:fe_int} to the dataset, varying $E(B-V)$ between 0.04 and 0.11 in 0.005~mag steps. For each value, the number of stars that fall strictly inside the MS (i.e., $\beta=0$) was recorded. The result is shown in Fig.~\ref{fig:red_cal}, where the reference reddening from \citet{Schlegel1998} is identified with a round marker. There are fluctuations in the upper region of the plot due to the discrete nature of the star count, but a plateau is evident for $E(B-V) \in [0.055,0.085]$. The absence of a definite peak could be, at least in part, a side effect of the irregular internal reddening. The median calibrated color excess is $E(B-V)_{cal}=0.07$~mag (hereafter, the subscript cal denotes values derived from the reddening calibration).

\section{Results and discussion}
\label{sec:results}

\subsection{Property identification}
\label{sec:prop_id}
\begin{figure}
	\includegraphics[width=\columnwidth]{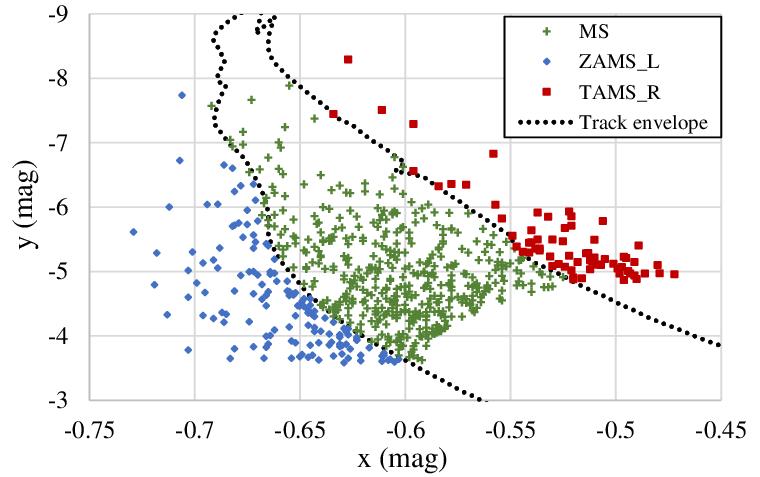}
	\caption{Identified objects: inside MS (green crosses), left of ZAMS (blue diamonds), right of TAMS (red squares). Dotted line marks envelope of MS tracks. $E(B-V)_{cal}=0.07$, M>8~M$_\odot$, $\beta=2$, [Fe/H]=$-1$, $V/V_{crit} = 0.4$, $\mu^0 = 25.63$.}
	\label{fig:final_cmd}
\end{figure}

From 9412 objects in the photometry dataset, using the calibrated reddening $E(B-V)_{cal}=0.07$ and setting $\beta=2$ (i.e., maximum distance to the MS below two standard deviations of color) yields 655 stars above 8~M$_\odot$ (Fig.~\ref{fig:final_cmd}). There are 448 objects strictly inside the main sequence tracks (MS, green crosses), 138 to the left of the ZAMS (ZAMS\_L, blue diamonds) and 69 to the right of the TAMS (TAMS\_R, red squares). For reference, the first and last points of all tracks have been connected with dotted black lines to outline the MS. Note that, even though the reddening calibration ensures that the observed CMD is centered on the MS, there are twice as many ZAMS\_L stars than TAMS\_R objects. This occurs because $dy/dx$ is positive along a track. Therefore, the ZAMS\_L group includes fainter objects. This asymmetry does not affect the reddening calibration, which uses a limiting magnitude instead of a minimum mass.

The variation of the star count with $\beta$ (the tolerance measured in standard deviations of color) is shown in Table~\ref{tab:nstar_beta}. Almost 90 per cent of the objects identified lie within $1 \sigma_x$ of the MS tracks.

\begin{table}
	\centering
	\caption{Cumulative fraction of identified stars vs. color tolerance ($\beta$).}
	\label{tab:nstar_beta}
	\begin{tabular}{lccccc}
		\hline
		$\beta$ & 0 & 0.5 & 1 & 1.5 & 2\\
		\hline
		Star count & 448 & 532 & 586 & 623 & 655\\
		Fraction (\%) & 68 & 81 & 89 & 95 & 100\\	
		\hline
	\end{tabular}
\end{table}

\subsection{Mass uncertainty and relation to photometric quality}
\label{sec:munc_qual}
\begin{figure}
	\includegraphics[width=\columnwidth]{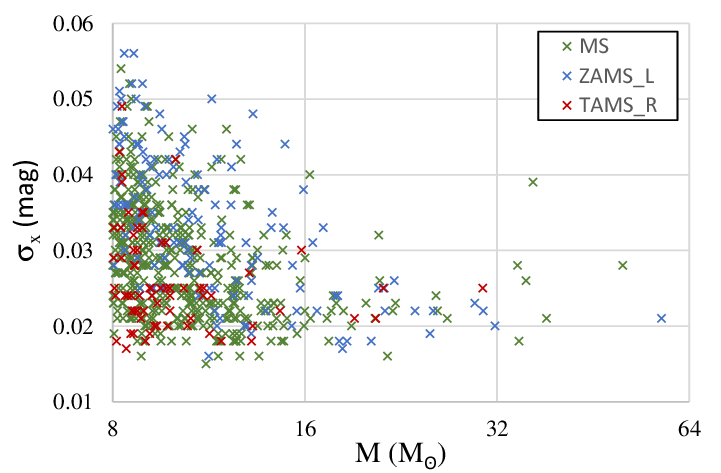}
	\caption{Color uncertainty vs. initial mass. Green markers: points inside MS, Blue: left of ZAMS, Red: right of TAMS. $E(B-V)_{cal}=0.07$, M>8~M$_\odot$, $\beta=2$, [Fe/H]=$-1$, $V/V_{crit} = 0.4$, $\mu^0 = 25.63$.}
	\label{fig:sx_m}
\end{figure}

\begin{figure}
	\includegraphics[width=\columnwidth]{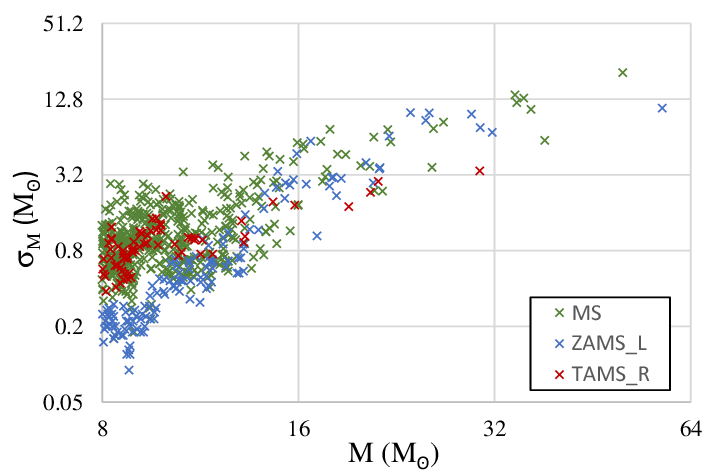}
	\caption{Mass uncertainty vs. initial mass. Green markers: points inside MS, Blue: left of ZAMS, Red: right of TAMS. $E(B-V)_{cal}=0.07$, M>8~M$_\odot$, $\beta=2$, [Fe/H]=$-1$, $V/V_{crit} = 0.4$, $\mu^0 = 25.63$.}
	\label{fig:sm_m}
\end{figure}

Plotting the color uncertainty against the initial stellar mass (Fig.~\ref{fig:sx_m}) shows that, as estimated in Sec.~\ref{sec:lim_mag}, $\sigma_x<0.05$~mag for the vast majority of massive stars. The photometric error of objects found left of the ZAMS (blue markers) is slightly larger than those inside the MS (green) and right of the TAMS (red). The reason is that, due to the slope of the tracks, objects of a given mass are fainter on the left side of the CMD. Objects above 18 Solar masses have color errors close to 0.025~mag. However, the increased photometric accuracy of the most massive stars does not translate into a better mass estimate (Fig.~\ref{fig:sm_m}) due to color degeneracy. In fact, their uncertainty becomes roughly proportional to the mass ($\sigma_M \sim 0.3M$).

\begin{figure}
	\includegraphics[width=\columnwidth]{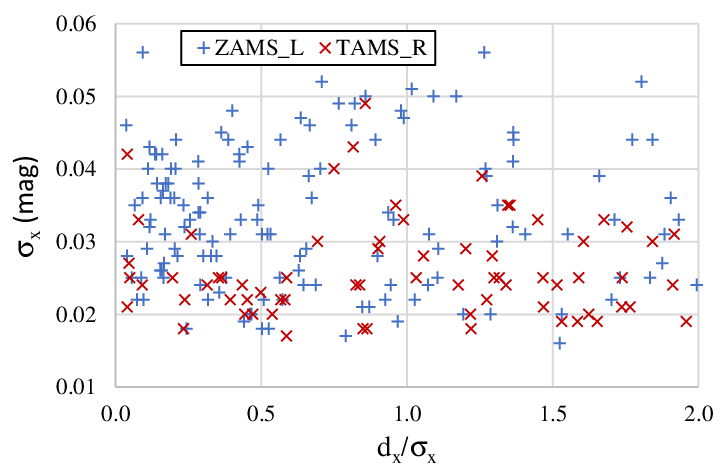}
	\caption{Color error vs. relative distance to MS track envelope. Blue markers: stars left of ZAMS, Red: objects right of TAMS. $E(B-V)_{cal}=0.07$, M>8~M$_\odot$, [Fe/H]=$-1$, $V/V_{crit} = 0.4$, $\mu^0 = 25.63$.}
	\label{fig:sx_dxsx}
\end{figure}

\begin{figure}
	\includegraphics[width=\columnwidth]{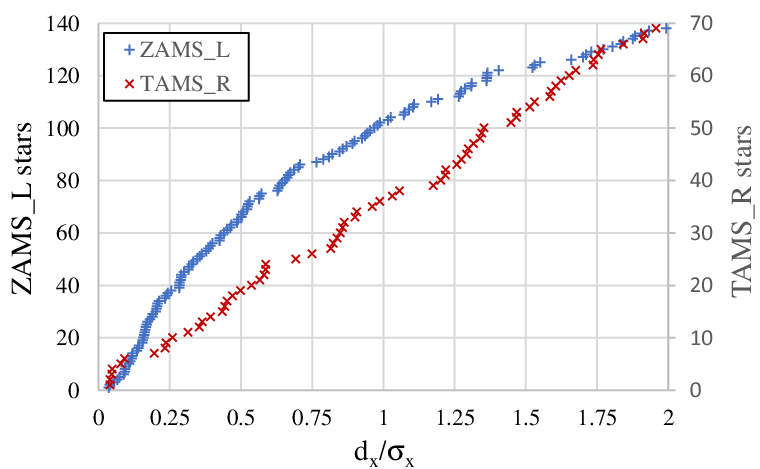}
	\caption{Cumulative number of stars vs. relative distance to MS track envelope. Blue markers: stars left of ZAMS, Red: objects right of TAMS. $E(B-V)_{cal}=0.07$, M>8~M$_\odot$, [Fe/H]=$-1$, $V/V_{crit} = 0.4$, $\mu^0 = 25.63$.}
	\label{fig:nstar_dxsx}
\end{figure}

It is interesting to examine the distribution of photometry errors and distances to the MS tracks for the objects identified outside of the envelope. Figure~\ref{fig:sx_dxsx} plots the color error against the relative distance ($d_x/\sigma_x$). There is no clear correlation between the quality of the photometry and the distance to the tracks. The majority of points lie close to the edge of the track set (in agreement with Table~\ref{tab:nstar_beta}), irrespective of $\sigma_x.$ This suggests that the fluctuations of internal reddening are moderate compared to the photometric uncertainty for the majority of objects. Otherwise, the distribution of relative distances would be more uniform. Figure~\ref{fig:nstar_dxsx} displays the cumulative frequency of stars against the relative distance. For the ZAMS\_L subset it is clear that the stars are clustered close to the track envelope. However, the TAMS\_R objects appear to be distributed much more uniformly. This might be an effect of the loss of accuracy of the stellar evolution models for \mbox{low-metallicity} stars. As explained before, they predict a large spurious gap between the MS and CHeB stages \citep{Cignoni2018}. This creates a gray zone where it is not possible to determine reliably if a star is undergoing helium burning. It may be the case that the actual MS extends further to the red end of the CMD than the tracks predict. Alternatively, some of the stars in the TAMS\_R subset may actually be in the CHeB stage. If the uniform distribution of distances for TAMS\_R objects were the result of substantial irregular internal reddening, the same effect would be expected for the ZAMS\_L stars. Given that this is not the case, the unreliability of the stellar evolution model seems a more plausible explanation.

\begin{figure}
	\includegraphics[width=\columnwidth]{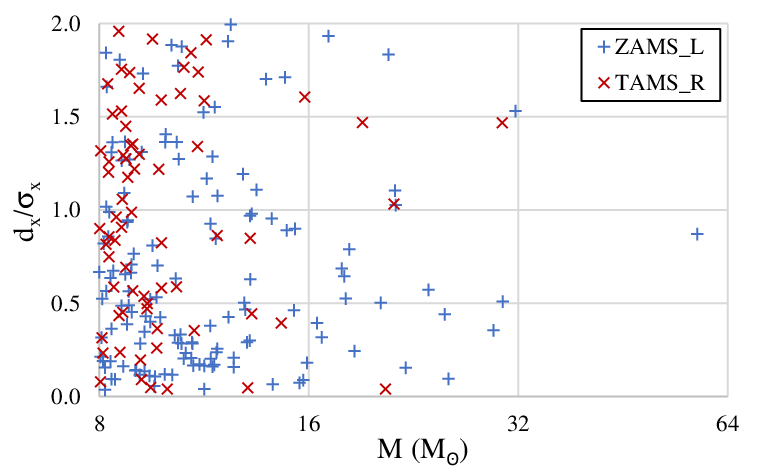}
	\caption{Relative distance to track envelope vs. stellar mass. Blue markers: stars left of ZAMS, Red: objects right of TAMS. $E(B-V)_{cal}=0.07$, M>8~M$_\odot$, [Fe/H]=$-1$, $V/V_{crit} = 0.4$, $\mu^0 = 25.63$.}
	\label{fig:dxsx_m}
\end{figure}

In Fig.~\ref{fig:dxsx_m} the variation of the relative distance against the stellar mass is shown. Heavier stars are brighter and therefore have smaller $\sigma_x$. If the internal reddening were appreciable for a large fraction of stars, the relative distance would increase for the most massive stars. No such trend is apparent in the graph. Thus, the results are compatible with the observations of Lerenzo et al. (2025), who found only a small fraction of stars affected by large internal color excess.

\subsection{The catalog of massive dwarf stars in Sextans A}
\label{sec:catalog}
\begin{figure}
	\includegraphics[width=\columnwidth]{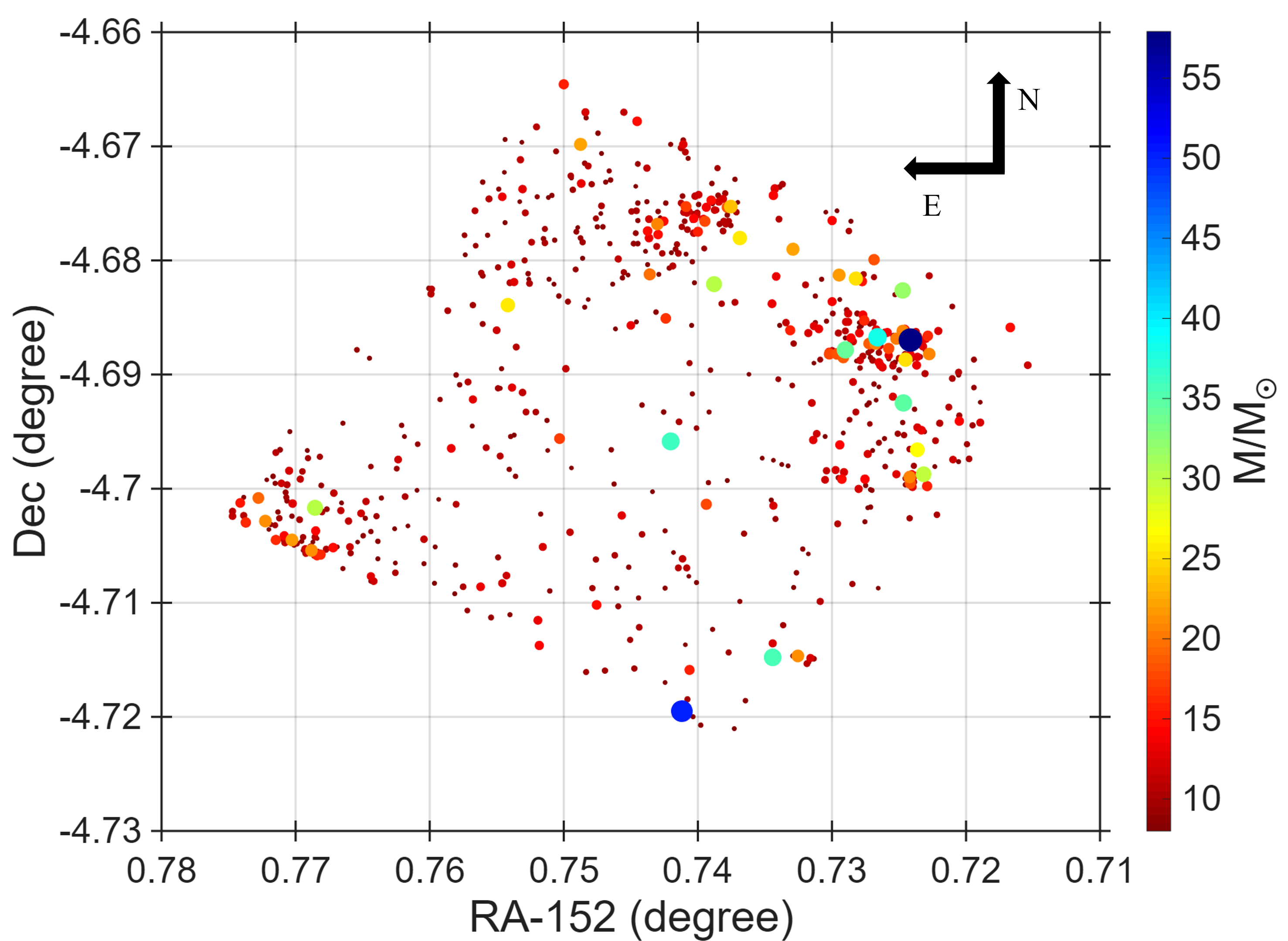}
	\caption{Positions of massive stars (M>8~M$_\odot$) in the catalog. Marker size and color indicate initial mass}
	\label{fig:mass_map}
\end{figure}

The positions of the 655 MS stars in our catalog are shown in Fig.~\ref{fig:mass_map}. The most massive stars are clustered near the corners of the field, matching to the four distinct zones of active star formation reported in the literature \citep{Garcia2019,Lorenzo2022,Lorenzo2025}.

The catalog contains the key stellar properties, including astrometry, photometry, initial mass, age, effective temperature and surface gravity, as well as associated uncertainties (see Table~\ref{tab:cat_key}). Table~\ref{tab:catalog} presents the results for the ten most massive objects. The complete catalog of 655 stars is available in the additional material.

To ensure that the catalog does not contain field objects, it was cross-matched with the Gaia EDR3 database, which is the reference for HST astrometry \citep{Mack2022}. Setting a tolerance of 100 mas for the search, only 8 matches are found (Table~\ref{tab:gaia_stars}), because most stars in Sextans A are too faint to be mapped by Gaia. For all matches, the difference in astrometry between HST and Gaia is below 10 mas, making the identification very reliable. The 8 objects have either no parallax and proper motion estimate, or the values are compatible with zero (i.e., close to the measurement uncertainty). Therefore, it is unlikely that they are foreground stars. 

\begin{table}
	\centering
	\caption{Key to Table~\ref{tab:catalog}.}
	\label{tab:cat_key}
	\begin{tabular}{ll}
		\hline
		Column & Value\\
		\hline
		IDN & Unique serial identifier\\
		Zone & Region of the CMD (ZAMS\_L/MS/TAMS\_R)\\
		M & ZAMS mass (M$_\circleddot$)\\
		$\sigma_\text{M}$ & Uncertainty of M\\
		RA, Dec & Right ascension/declination in ICRS (degree)\\
		F275, F336 & Apparent magnitude in F275W/F336W filter (Vega system)\\
		$\sigma_\text{F275}$, $\sigma_\text{F336}$ & Uncertainty of F275/F336\\
		x, y & Coordinates in CMD (mag)\\
		d$_\text{x}$/$\sigma_\text{x}$ & Distance to MS tracks in standard deviations of color\\
		Age & Stellar age (Ma)\\
		log(T) & Decimal logarithm of effective temperature (K)\\
		log(g) & Decimal logarithm of surface gravity (Gal)\\
		$\sigma_\text{Age}$, $\sigma_\text{logT}$, $\sigma_\text{logg}$ & Uncertainty of Age/log(T)/log(g)\\
		\hline
	\end{tabular}
\end{table}

\begin{landscape}
\begin{table}
	\centering
	\caption{Sample of the catalog of massive main sequence stars in Sextans A. See Table~\ref{tab:cat_key} for the key to the columns.}
	\label{tab:catalog}
	\begin{tabular}{lcccccccccccccccccc}
		\hline
		IDN & Zone & M & $\sigma_\text{M}$ & RA & Dec & F275& F336 & $\sigma_\text{F275}$ & $\sigma_\text{F336}$ & x & y & d$_\text{x}$/$\sigma_\text{x}$ & Age & log(T) & log(g) & $\sigma_\text{Age}$ & $\sigma_\text{logT}$ & $\sigma_\text{logg}$ \\
		\hline
		1 & ZAMS\_L & 57.9 & 10.9 & 152.7241585 & -4.6869743 & 18.277 & 18.910 & 0.021 & 0.018 & -0.706 & -7.737 & 0.87 & 0.0 & 4.73 & 4.37 & 2.4 & 0.10 & 0.33 \\
		2 & MS     & 50.4 & 20.8 & 152.7411774 & -4.7195234 & 18.441 & 19.060 & 0.028 & 0.025 & -0.692 & -7.573 & 0.02 & 0.6 & 4.70 & 4.31 & 3.9 & 0.10 & 0.36 \\
		3 & MS     & 38.2 & 6.0 & 152.7266164 & -4.6867412 & 18.971 & 19.581 & 0.021 & 0.019 & -0.683 & -7.043 & - & 0.1 & 4.67 & 4.37 & 0.2 & 0.05 & 0.13 \\
		4 & MS     & 36.4 & 10.6 & 152.7420465 & -4.6958470 & 19.069 & 19.678 & 0.039 & 0.028 & -0.682 & -6.945 & - & 0.0 & 4.67 & 4.39 & 0.5 & 0.10 & 0.24 \\
		5 & MS     & 35.5 & 13.0 & 152.7344452 & -4.7147598 & 18.347 & 18.947 & 0.022 & 0.026 & -0.673 & -7.667 & - & 4.2 & 4.64 & 4.04 & 3.3 & 0.10 & 0.36 \\
		6 & MS     & 34.6 & 12.0 & 152.7246869 & -4.6924857 & 18.846 & 19.450 & 0.016 & 0.018 & -0.677 & -7.168 & - & 2.1 & 4.66 & 4.26 & 4.2 & 0.06 & 0.23 \\
		7 & MS     & 34.4 & 13.9 & 152.7289993 & -4.6878345 & 19.042 & 19.646 & 0.028 & 0.019 & -0.677 & -6.972 & - & 0.7 & 4.66 & 4.33 & 5.6 & 0.11 & 0.39 \\
		8 & ZAMS\_L & 31.8 & 7.0 & 152.7247133 & -4.6826297 & 19.293 & 19.927 & 0.018 & 0.020 & -0.707 & -6.721 & 1.53 & 0.0 & 4.66 & 4.40 & 0.6 & 0.08 & 0.20 \\
		9 & ZAMS\_L & 30.4 & 7.6 & 152.7685622 & -4.7016954 & 19.362 & 19.975 & 0.022 & 0.021 & -0.686 & -6.652 & 0.51 & 0.0 & 4.65 & 4.40 & 0.6 & 0.09 & 0.22 \\
		10 & TAMS\_R & 30.4 & 3.5 & 152.7387820 & -4.6820961 & 17.725 & 18.279 & 0.018 & 0.025 & -0.627 & -8.289 & 1.47 & 6.9 & 4.55 & 3.57 & 0.8 & 0.03 & 0.11 \\
		\hline
	\end{tabular}
\end{table}
\begin{table}
	\centering
	\caption{Objects with a counterpart in Gaia EDR3.}
	\label{tab:gaia_stars}
	\begin{tabular}{lccccc}
		\hline
		IDN & Separation (mas) & Gaia EDR3 identifier & RA (degree) & Dec (degree) & G-band magnitude\\
		\hline
		1 & 2.6	& 3779988382066051840 & 152.7241592 & -4.6869743 & 20.69\\
		5 & 7.3	& 3779988077121167616 & 152.7344466 & -4.7147613 & 20.66\\
		10 & 2.7 & 3779988558157500672 & 152.7387827 & -4.6820959 & 19.93\\
		14 & 4.2 & 3779988764316004480 & 152.7368555 & -4.6780290 & 20.47\\
		21 & 3.2 & 3779987733524193152 & 152.7702547 & -4.7044620 & 20.89\\
		25 & 4.9 & 3779988798675667072 & 152.7430337 & -4.6768353 & 20.95\\
		32 & 3.3 & 3779987767883386624 & 152.7727766 & -4.7008284 & 20.81\\
		191 & 5.8 & 3779988558157502336 & 152.7334290 & -4.6854748 & 20.80\\
		\hline
	\end{tabular}
\end{table}
\end{landscape}

\subsection{Overlap with a spectroscopic catalog of OB stars in Sextans A}
\label{sec:hst_lor}
\begin{figure}
	\includegraphics[width=\columnwidth]{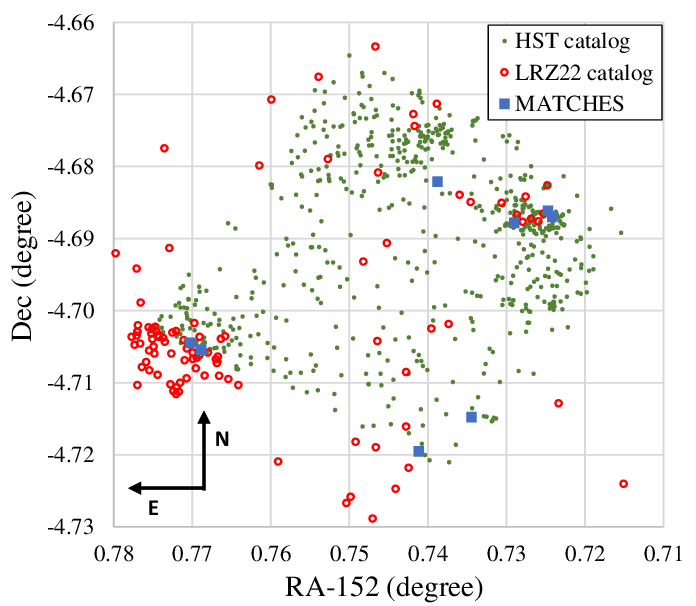}
	\caption{Spatial overlap of massive stars in our catalog (green dots) and LRZ22 (red circles). Blue squares indicate shared objects.}
	\label{fig:lrz_hst_overlap}
\end{figure}

\citet{Lorenzo2022} built a census of massive stars in Sextans A from ground-based observations. It contains the detailed spectroscopic classification (type, subtype, luminosity class and peculiar features) of 118 OB stars. This dataset of 118 objects shall be denoted as LRZ22 hereafter. The reference also includes a list of objects with poor-quality data which could not be classified precisely, they are addressed separately in Sec.~\ref{sec:match_lowq}. To illustrate how the present work complements existing literature, we cross-matched the two datasets to determine the level of commonality. As shown in Fig.~\ref{fig:lrz_hst_overlap}, LRZ22 covers a wider sky area than the HST field (it is actually larger than what is depicted in the image).

\subsubsection{Astrometric calibration}
\label{sec:ast_cal}
In order to identify which objects belong to both catalogs, an astrometric correction was necessary. A preliminary search revealed differences in position above 0.1~arcsec even for the closest matches. This is likely due to inconsistent astrometry sources. HST positions are taken from the Hubble data reduction pipeline, which is referenced to Gaia eDR3 \citep{Mack2022}. On the other hand, LRZ22 uses astrometry from \citet{Massey2007}, calibrated with the USNO-B1.0 catalog \citep{Monet2003}.

To improve the reliability of the identification, we matched all stars in the HST field above 6~M$_\circleddot$, including the MS and CHeB evolutionary stages, against LRZ22. While some of these objects are not part of the final catalog, they are useful as reference points to anchor the photometry. The tolerance for finding matches was set to 0.4~arcsec. Objects closer than this distance would not be adequately resolved with ground instruments, leading to errors.

There are 41 objects in LRZ22 with at least one counterpart in our dataset within 0.4~arcsec. Furthermore, 18 of them have more than one nearby companion (i.e., several stars in our dataset separated less than 0.4~arcsec from the coordinates given in LRZ22). This highlights the higher susceptibility to crowding of the ground observations. To resolve ambiguities, the U, B and V magnitudes of the HST stars were predicted using synthetic photometry from the stellar evolution model, and compared to the values given in LRZ22.

All the matches with large discrepancies in photometry (>0.4~mag in any channel) were discarded. Another two objects were considered unreliable because the mass estimates were incompatible with the spectral classification. Finally, 3 objects with multiple matches were pruned because the candidate with the best predicted ground photometry was not the closest to the target coordinates. Given the possibility of an incorrect identification due to the presence of companions not resolved from the ground, they were considered unreliable.

After filtering, 24 stars remained. Computing the median and standard deviation of the differences in right ascension and declination for this subset gives
\begin{equation}
	\begin{array}{l}
		\text{RA}_{HST}-\text{RA}_{LRZ22} = -0.143 \pm 0.051 \; \text{arcsec},\\
		\text{Dec}_{HST}-\text{Dec}_{LRZ22} = -0.090 \pm 0.037 \; \text{arcsec},
	\end{array}
	\label{eq:ast_cor}
\end{equation}
which is a strong indicator of a systematic difference in the astrometry.

\subsubsection{Common objects and compatibility with LRZ22 classification}
\label{sec:match_lrz22}
Removing the median astrometric difference and restricting the search to MS stars above 8 Solar masses, with a tolerance of 0.1~arcsec \citep[slightly above the 0.08~arcsec PSF of the WFC3,][]{WFC3_IHB}, yields 8 matches (Table~\ref{tab:matches}). Thus, only 1.2 per cent on the stars in our census are included in LRZ22. The objects in our catalog are excellent candidates for followup spectroscopic observations, which would shed further light in the hydrogen-burning phase of massive stars.

\begin{table}
	\centering
	\caption{Massive stars with a counterpart in LRZ22. IDN/ID$_{\text{LRZ22}}$: identifier in our catalog/LRZ22, Sep: separation after astrometric correction (mas), Zone: region of the CMD, SpT: spectral classification (LRZ22), M/M$_{\text{SpT}}$: initial mass from CMD/Spectral type ($M_\odot$). M$_{\text{SpT}}$ taken from \citet{Weidner2010} for SMC metallicity (0.2~Z$_\odot$). {\bf NOTE}: LRZ22 signals s004 as a possible binary, due to changes in spectral features between observing runs.}
	\label{tab:matches}
\begin{tabular}{lcccccc}
	\hline
	IDN & ID$_{\text{LRZ22}}$ & Sep & Zone & SpT & M & M$_{\text{SpT}}$ \\
	\hline
	1 & s014 & 40 & ZAMS\_L & O7.5 III((f)) & 47-69 & 29-37 \\
	2 & s004 & 32 & MS & O5 III & 29-71 & 39-53 \\
	5 & s016 & 71 & MS & O8 II & 23-49 & 28-60 \\
	7 & s080* & 86 & MS & B1 III & 20-48 &  \\
	10 & s061 & 70 & TAMS\_R & B0 I & 26-34 &  \\
	21 & s046* & 26 & MS & O9.5 V & 20-24 & 13-23 \\
	24 & s054 & 9 & TAMS\_R & O9.7 III & 18-24 & 20-31 \\
	28 & s087 & 50 & TAMS\_R & B2 III + O? & 19-23 &  \\
	\hline
\end{tabular}
\end{table}

All objects in Table~\ref{tab:matches}, except for IDN 21 have been assigned luminosity classes I, II or III in LRZ22. This may seem to contradict our catalog, where they are classified as MS stars. Two reasons could explain this apparent discrepancy:
\begin{itemize}
	\item As noted in Sec.~\ref{sec:prep_track}, in \mbox{low-metallicity} environments the stellar evolution models predict an unrealistically wide gap between the MS and CHeB tracks. Thus, it would not be surprising for stars found to the right of TAMS to be undergoing helium burning. This affects IDN 10, 24 and 28.
	
	\item Very massive stars can leave luminosity class V before they finish the hydrogen burning stage \citep[][Appendix B]{Weidner2010}. This is applicable to IDN 1, 2, 5 and 7.
\end{itemize}
Additionally, s004 (which matches IDN 2) is marked in LRZ22 as a possible binary, due to changes in spectral features between observing runs.

To provide an approximate quantitative comparison between both catalogs, column M$_{\text{SpT}}$ of Table~\ref{tab:matches} contains the initial mass estimated by \citet{Weidner2010} as a function of the spectroscopic classification. The values given are for the SMC metallicity (0.2~Z$_\odot$), the lowest available in the reference. Unfortunately, no data is given for B-type stars. Furthermore, only luminosity classes I, III and V are provided. The mass range indicated for O8 II encompasses the intervals for O8 I and O8 III. \citet{Weidner2010} does not provide values for O9.7 stars, the mass listed for IDN 24 corresponds to subtype O9.5 III.

Despite all the caveats above, the two mass estimates are compatible for all objects except IDN 1. Evolutionary models often predict larger masses than the spectroscopic analysis of stellar atmospheres \citep{Herrero1992,Mokiem2007,Weidner2010}, sometimes by a factor as large as 2. Thus, the discrepancy is not surprising. That being said, IDN 1 is in a region of the CMD with quasi-overlapping and highly-distorted stellar tracks. Due to the near-degeneracy of the tracks, the estimated properties are extremely sensitive to color index errors, which are probably at play because the star is detected outside the MS (ZAMS\_L region). Moreover, the extinction map of \citet{Lorenzo2025} gives $E(B-V) \sim 0.15$ in the immediate vicinity of this object\footnote{The reference does not provide a table with numerical values, only a color map. Therefore, the authors do not know the precise extinction calculated by \citet{Lorenzo2025}. However, the star is clearly located in a region of elevated reddening.}. Because the global reddening correction cannot compensate for irregular internal extinction, the results for IDN 1 should be treated with caution.

\subsubsection{A common object with imprecise spectroscopic classification}
\label{sec:match_lowq}
\cite{Lorenzo2022} includes a list of blue stars that could not be assigned a spectral subtype, due to poor S/N. Matching this set with our catalog yielded a single hit (Table~\ref{tab:match_lowq}). This information could prove valuable to improve the classification of the object.

\begin{table}
	\centering
	\caption{A massive star with imprecise spectral classification in \citet{Lorenzo2022}. IDN/ID$_{\text{LRZ22}}$: identifier in our catalog/LRZ22, Sep: separation after astrometric correction (mas), Zone: region of the CMD, SpT: spectral classification (LRZ22), M: initial mass from CMD (M$_\odot$).}
	\label{tab:match_lowq}
	\begin{tabular}{lccccc}
		\hline
		IDN & ID$_{\text{LRZ22}}$ & Sep & Zone & SpT & M\\
		\hline
		32 & s140* & 65 & TAMS\_R & B III & 17-21\\
		\hline
	\end{tabular}
\end{table}

\section{Conclusions}
\label{sec:conclusions}
We developed a catalog of massive (M>8~M$_\odot$) stars in the \mbox{low-metallicity} dwarf irregular galaxy Sextans A. HST UV photometry in the $F275W$ and $F336W$ wide-band filters was combined into a CMD. The stellar properties were estimated comparing the observed CMD with synthetic photometry from the MIST stellar evolution library. A finite element interpolation gives the estimated properties as well as their sensitivities to errors in photometry.

While the methodology presented is applicable to all luminosity classes, giants have been excluded from the catalog due to the degraded accuracy in the CHeB stage. The issue is caused by the limited reliability of the evolutionary models in \mbox{low-metallicity} environments, as well as increased degeneracies due to intersections between tracks.

The results are highly sensitive to uncertainties that affect the color index (e.g., foreground and internal reddening, zero-point errors of the photometry, inaccuracies of the stellar evolution models).  Notably, dust maps based on far IR emission, which underpin the reference foreground extinction estimate $E(B-V)_{SWF}=0.044$ \citep{Schlegel1998}, are unreliable for external star-forming galaxies. The internal dust emission can distort the measurement of the Milky Way contribution. With an alternative technique, \citet{Massey2007} estimated $E(B-V)=0.05\pm0.05$. The uncertainty is comparable to the width of the MS in the CMD, highlighting the enormous sensitivity of the estimated properties to this parameter. This issue is compounded by the fact that the photometric color uncertainty is also of the same order.

To compensate for these limitations, a heuristic reddening calibration was developed. It is based on the assumption that the best reddening correction results in the largest number of stars identified inside the MS. The color excess in the visible range is taken as a free parameter, from which the UV extinctions are derived using the dust calibration curves of \citet{Schlafly2011}. The star count inside the MS tracks is plotted against $E(B-V)$, and the maximum of the curve determines the calibrated reddening. The result is $E(B-V)_{cal}=0.07\pm0.015$, higher than reference value from \citet{Schlegel1998}, but comparable considering the uncertainties involved. It must not be taken as a physical reddening. Instead, it is a correction that yields the most plausible relative position of the stars and the evolutionary tracks.

Using the calibrated reddening, 655 stars above 8~M$_\odot$ were identified. This includes 448 directly inside the MS tracks, 138 left of the ZAMS and 69 right of the TAMS. The stars outside of the track envelope have been retained because their separation is less than twice the color uncertainty of the photometry. Thus, is reasonable to assume that they are MS objects displaced due to photometric errors. An analysis of the distribution of relative distances (i.e., measured in standard deviations of color) to the MS shows that the majority of objects are clustered near the tracks. This suggest that the primary driver of the separation is the photometric uncertainty, because irregular internal reddening should lead to a more uniform distribution. This is reaffirmed by the fact that the distribution does not seem to depend on the mass of the objects, which affects the color uncertainty. If the internal reddening were significant, it should have a larger relative impact on the heaviest stars. However, this effect is not observed. The results are compatible with limited internal reddening, which agrees with the observations of \citet{Lorenzo2025}.

In an effort to identify foreground objects, the catalog was cross-matched against the Gaia EDR3 database. It resulted in only 8 hits because MS stars are faint in the visible spectrum, and most of them fall outside the sensitivity of Gaia. All matches were reliable (astrometric separation below 10 mas) and without significant parallax and proper motion. Therefore, they are likely members of Sextans A.

Finally, our dataset was matched with the \citet{Lorenzo2022} spectroscopic catalog (LRZ22). Only 8 reliable hits were found, less than 2 per cent of the objects in our catalog. Part of the reason is that the two datasets focus on different object types. Our census is exclusively for MS stars (although, it might include some CHeB stars, due to the reduced fidelity of stellar evolution models in \mbox{low-metallicity} environments). On the other hand, $\sim75$ per cent of the stars in LRZ22 are giants (luminosity classes I, II and III), because they are more suitable targets for ground telescopes. Due to the minimal overlap between both datasets, our catalog provides a valuable list of targets for future spectroscopic studies. They would yield further insight into the early evolutionary stages of \mbox{low-metallicity} massive stars, complementing existing observations.

\section*{Acknowledgements}
This research has been supported by Polar Research Center (PRC), Khalifa University of Science and Technology (KU-PRC). EF received support from the 
Spanish Ministry of Science and Innovation through grant PID2021-123968NB-I00. AA acknowledges the hospitality of Khalifa University of Science and Technology. AA has been partially supported by grants PID2020-115981GB-I00 and PID2023-150319NB-C22 (Spanish Ministry of Science and Innovation). GB acknowledges support from the Agencia Estatal de Investigación del Ministerio de Ciencia, Innovación y Universidades (MCIU/AEI) under grant EN LA FRONTERA DE LA ARQUEOLOGÍA GALÁCTICA: EVOLUCIÓN DE LA MATERIA LUMINOSA Y OSCURA DE LA VÍA LÁCTEA Y
LAS GALAXIAS ENANAS DEL GRUPO LOCAL EN LA ERA DE GAIA (FOGALERA) and the European Regional Development Fund (ERDF) with reference PID2023-150319NB-C21 / 10.13039/501100011033.\\
The authors express their gratitude to Drs. Artemio Herrero and Santi Cassisi for their advice on the internal extinction of massive stars.\\
Facility: NASA/ESA Hubble Space Telescope WFC3/UVIS\\
Software: WFC3/UVIS standard calibration pipeline, IRAF \citep{Stetson1987}

\section*{Data Availability}
The complete catalog of massive stars in Sextans A is provided in the additional material, both in formatted and machine readable versions.

\bibliographystyle{mnras}
\bibliography{References_V01}


\bsp	
\label{lastpage}
\end{document}